\documentstyle{lamuphys}
\newcommand{\T}{\textstyle}
\renewcommand{\D}{\displaystyle}
\makeatletter
\let\chapter\hid@chapter
\makeatother
\begin{document}
\pagenumbering{arabic}
\title{
Fermions and Condensates on the Light-Front
}

\author{M.\,Burkardt }
\institute{  }
\institute{Department of Physics,
New Mexico State University, Box 30001 Dept.3D,
Las Cruces, NM 88003-0001,U.S.A.}

\maketitle

\begin{abstract}
Light-Front quantization is one of the most promising and physical 
tools towards studying deep inelastic scattering on the basis of 
quark gluon degrees of freedom. The simplified vacuum structure 
(nontrivial vacuum effects can only appear in zero-mode degrees of 
freedom) and the physical basis allows for a description of hadrons 
that stays close to intuition. I am reviewing
recent progress in understanding the deep connection between 
renormalization of light-front Hamiltonians, effective 
light-front Hamiltonians and nontrivial vacuum condensates.
\footnote{Lecture Notes, based on three lectures given
at the ``School on Light-Front Quantization and Non-Perturbative Dynamics 
- Theory of Hadrons and Light-Front QCD'', IITAP, Ames, IA, May 1996.}
\end{abstract}
\section{Advantages of Light-Front Coordinates}
\begin{figure}
\unitlength1.cm
\begin{picture}(14,6)(-1.5,-8.)
\includegraphics{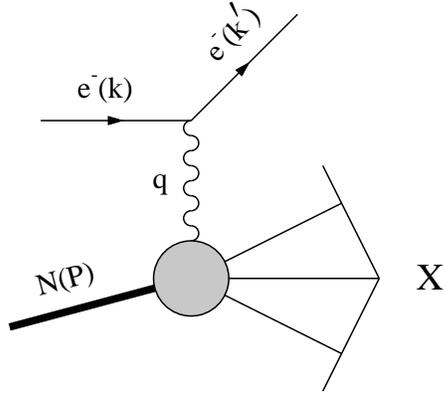}
\end{picture}
\caption{Inclusive process $e^-+N \rightarrow e^{-\prime}+X$,
where $X$ is an unidentified hadronic state.
}
\label{fig:dis}
\end{figure}

\begin{figure}
\unitlength1.cm
\begin{picture}(15,6)(1.4,-9)
\includegraphics{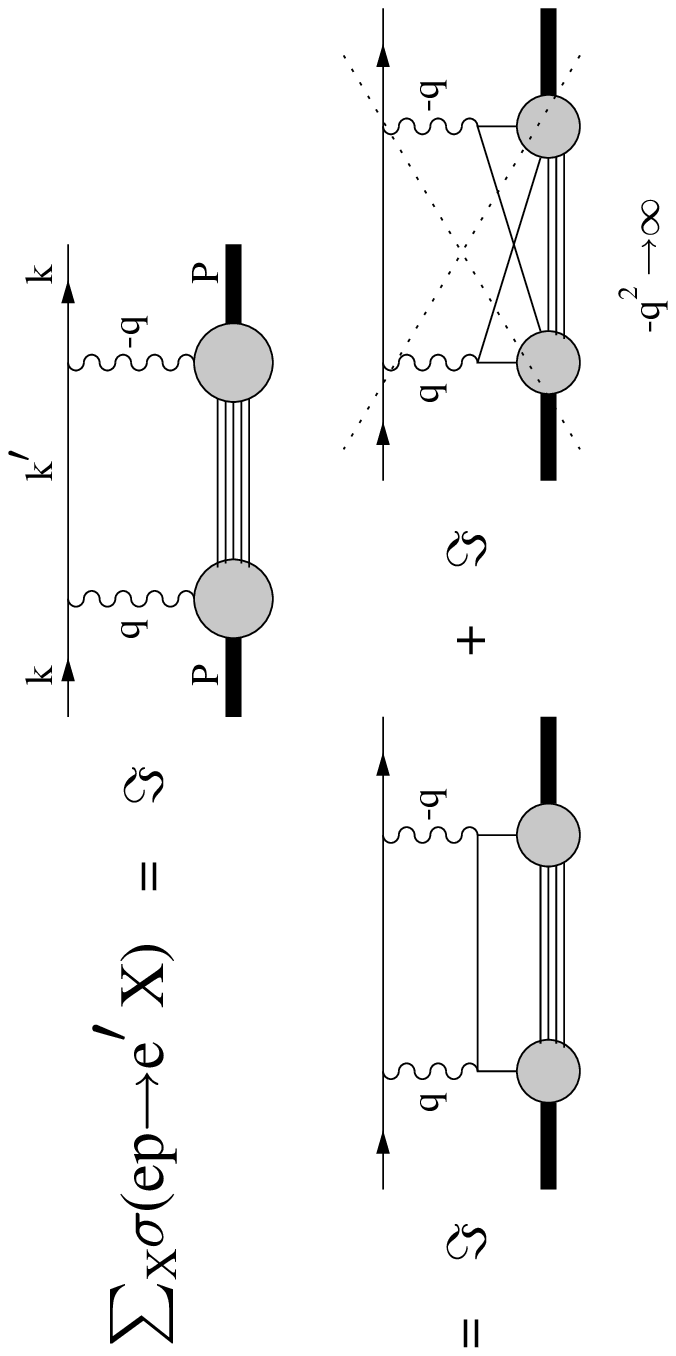}
\end{picture}
\caption{Inclusive lepton nucleon cross section expressed
in terms of the imaginary part of the forward Compton amplitude.
For $Q^2=-q^2\rightarrow \infty$ only the `handbag diagram'
(both photons couple to the same quark) survives. The `crossed
diagram' (the two photons couple to different quarks) is
suppressed because of wavefunction effects.
}
\label{fig:compt}
\end{figure}

Deep inelastic lepton-nucleon scattering (DIS) provides 
access to quark and gluon degrees of freedom in nucleons
and nuclei. In these experiments one shoots high energy 
leptons (e.g. electrons) at a hadronic target
(usually protons or nuclei) and measures the energy and
momentum transfer to the target by detecting the
final state lepton (Fig. \ref{fig:dis}).
In the most simple version of DIS,
the hadronic final state $X$ is not measured
(usually the nucleon is destroyed in these reactions and the
hadronic final state consists of many particles).
Because of the extremely large momentum transfer to the
target (typical momentum transfers in DIS experiments are
several $GeV/c$ or more), 
the inclusive cross sections are dominated by single
particle response functions along the light-cone.
To illustrate this let us use the optical theorem which
relates the differential lepton nucleon cross section 
to the imaginary part of the forward Compton amplitude
\cite{yn:qcd} (Fig.\ref{fig:compt}).
One finds
\begin{equation}
\frac{d^2\sigma}{d\Omega dE^\prime}
= \frac{\alpha^2}{q^4} \left( \frac{E^\prime}{E}
\right) l^{\mu \nu}
\frac{\Im T_{\mu \nu}}{2\pi}
\end{equation}
where $E, E^\prime$ are the energies of the initial and final 
lepton. Furthermore, \mbox{$l_{\mu\nu}=2k_\mu k^\prime_\nu + 2k_\nu k^\prime_\mu +
q^2g_{\mu \nu}$} is the leptonic tensor and 
\mbox{$q=k-k^\prime$}
is the four momentum transfer of the lepton on the
target. The hadronic
tensor
\begin{equation}
T_{\mu \nu}(P,q) = \frac{i}{2M_N} \sum_S\int \frac{d^4x}{2\pi}
e^{iq\cdot x}
\langle P,S |T\left(J_\mu(x) J_\nu(0)\right)|P,S \rangle
\label{eq:htensor}
\end{equation}
($S$ is the spin of the target proton)
contains all the information about the parton substructure
of the target proton.

In the Bjorken limit ($Q^2 \equiv -q^2 \rightarrow \infty$,
$P\cdot q \rightarrow \infty$, $x_{Bj} = Q^2/2P\cdot q$ fixed),
deep inelastic structure functions exhibit Bjorken scaling:
up to kinematic coefficients, the hadronic tensor
(\ref{eq:htensor}) depends only on $x_{Bj}$ but no longer
on $Q^2$ (within perturbatively calculable logarithmic
corrections). In order to understand this result, it is
convenient to introduce {\it light-front} variables
$a_\mp = a^\pm = \left( a^0 \pm a^3\right)/\sqrt{2}$ so that
the scalar product reads
$a \cdot b = a_+b^+ + a_- b^- -{\vec a}_\perp 
{\vec b}_\perp =a_+b_- + a_- b_+ -{\vec a}_\perp 
{\vec b}_\perp$. Furthermore let us choose a frame where 
${\vec q}_\perp =0$. The Bjorken limit corresponds to
$p^\mu$ and $q_-$ fixed, while $q_+\rightarrow \infty$. 
Bjorken scaling is equivalent to the statement that 
the structure functions become independent of $q_+$ in this limit (again up to trivial kinematic coefficients).
In this limit, the integrand in Eq.(\ref{eq:htensor}) contains the rapidly oscillating factor 
$\exp (iq_+x^+)$, which kills all contributions to the 
integral except those where the integrand is singular
\cite{jaffe}. Due to causality, the integrand must vanish 
for $x^2=2x^+x^- -{\vec x}^2_\perp <0$ and the current 
product is singular at $x^+=0$, ${\vec x}_\perp =0$.
The leading singularity can be obtained from the
operator product expansion by contracting two fermion
operators in the product
$T\left( J_\mu(x) J_\nu(0) \right) \equiv
T\left( \bar{\psi}(x) \gamma_\mu \psi(x)
\bar{\psi}(0) \gamma_\nu \psi(0) \right)$, yielding
a nonlocal term bilinear in the fermion field multiplying
a free (asymptotic freedom!) fermion propagator from $0$ to $x$
which gives rise to the abovementioned singularity structure
\cite{ch:gau}.
The $x^+={\vec x}_\perp =0$ dominance in the integral has two consequences.
First it explains Bjorken scaling, because $q_+$ enters the
hadronic tensor only via the term $x^+q_+$ in the exponent
and for $x^+=0$ the $q_+$ dependence drops out. Second,
and this is very important for practical calculations,
the parton distributions, i.e. the Bjorken scaled 
structure functions, can be expressed in terms
of correlation functions along the {\it light-front} 
space direction $x^-$.
For example, for the spin
averaged parton distribution one obtains
\begin{equation}
2P_-f(x_{Bj}) = \int \frac{dx^-}{2\pi} \langle P |
\bar{\psi}(0)\gamma_- \psi(x^-)|P \rangle \exp (iP_-x^- x_{Bj}),
\label{eq:parton}
\end{equation}

The physical origin of this result can be understood as
follows. Consider again the virtual forward Compton 
amplitude (Fig. \ref{fig:compt}).
In principle, the photons in the first and second interaction
in Fig. \ref{fig:compt} can couple to the same as well as to different
quarks in the target. However, the hadronic wavefunction can
only absorb momenta which are of the order of the
QCD-scale ($\Lambda_{QCD}\approx 200 MeV$).
Therefore, in the limit of large momentum transfer, only
such diagrams survive where the two photons in Fig. \ref{fig:compt}
couple to the same quark. All other diagrams have
large momenta flowing through the wavefunction or they
involve extra hard gluon exchanges which results in 
their suppression at large $Q^2$.
The large momentum transfer is also important because of 
asymptotic freedom. Since $\alpha_S(Q^2)\sim 1/\log \left(Q^2/\Lambda_{QCD}^2\right)$,
the running coupling constant of QCD,
goes to zero for large $Q^2$, all interactions of
the struck quark can be neglected and it propagates
essentially without interaction
between the two
photon-vertices. Furthermore, since the momentum
transfer is much larger than the masses of the quarks
in the target, the struck quark's propagation between
becomes ultra-relativistic, i.e. it moves
exceedingly close to the light cone $x^2=0$.
Due to the high-energy nature of the scattering, the
relativistic structure function is a LF correlation
\cite{rj:70,ji:com}. Already at this point it should
be clear that LF-coordinates play a distinguished
role in the analysis of DIS experiments ---
a point which will become much more obvious after
we have introduced some of the formal ideas of
LF quantization.

LF quantization is very similar to canonical equal 
time (ET) quantization \cite{di:49} (here we closely follow Ref. \cite{kent}). 
Both are Hamiltonian formulations of
field theory, where one specifies the fields on a
particular initial surface. The evolution of the fields
off the initial surface is determined by the 
Lagrangian equations of motion. The main difference 
is the choice of the initial surface, $x^0=0$ for
ET and $x^+=0$ for the LF respectively. 
In both frameworks states are expanded in terms of fields
(and their derivatives) on this surface. Therefore,
the same physical state may have very different
wavefunctions\footnote{By ``wavefunction'' we mean here
the collection of all Fock space amplitudes.}
in the ET and LF approaches because fields at $x^0=0$ 
provide a different basis for expanding a state than 
fields at $x^+=0$. The reason is that the microscopic 
degrees of freedom --- field amplitudes at $x^0=0$ 
versus field amplitudes at $x^+=0$ --- are in general
quite different from each other in the two formalisms. 

This has important consequences for the practical
calculation of parton distributions (\ref{eq:parton})
which are real time response functions in the equal
time formalism.
\footnote{The arguments of $\bar{\psi}$ and $\psi$ in 
Eq.(\protect\ref{eq:parton}) have different time components!}
In order to evaluate Eq.(\protect\ref{eq:parton}) one needs to
know not only the ground state wavefunction of
the target, but also matrix elements to excited states. 
In contrast, in the framework of LF quantization,
parton distributions are correlation functions at equal
LF-time $x^+$, i.e. 
{\it within} the initial surface $x^+=0$ and can thus
be expressed directly in terms of ground state
wavefunctions (As a reminder: ET wavefunctions and
LF wavefunctions are in general different objects).
In the LF framework, parton distributions $f(x_{Bj}$) can
be easily calculated and have a very simple physical 
interpretation as single particle momentum densities, 
where $x_{Bj}$ measures the fraction of the hadron's
momentum that is carried by the parton
\footnote{In DIS with non-relativistic kinematics
(e.g. thermal neutron scattering off liquid $^4$He)
one also observes scaling and the structure functions can be
expressed in terms of single particle response functions.
However, due to the different kinematics, non-relativistic
structure functions at large momentum transfer
are dominated by Fourier transforms of
equal time response functions, i.e. ordinary momentum 
distributions.}
\begin{equation}
x_{Bj} = \frac{p_-^{parton}}{P_-^{hadron}}.
\end{equation}

Although DIS is probably the most prominent example for 
practical applications of LF coordinates, they prove useful in many other places as well. For example, 
LF coordinates have been used in the context current algebra 
sum rules in particle physics \cite{fu:inf}.
Another prominent example is form factors, where moments
of the wave function along the LF determine the asymptotic
falloff at large momentum transfer \cite{br:lep}.
More recently, LF quantization found applications
in inclusive decays of heavy quarks \cite{bj:hq,mb:hq,wz:hq}.

From the purely theoretical point of view, various advantages
of LF quantization derive from properties of the ten generators
of the Poincar\'e group (translations $P^\mu$,
rotations ${\vec L}$ and boosts ${\vec K}$) \cite{di:49,kent}.
Those generators which leave the initial surface
invariant (${\vec P}$ and ${\vec L}$ for ET and
$P_-$, ${\vec P}_\perp$, $L_3$ and ${\vec K}$ for LF)
are ``simple'' in the sense that they have very simple
representations in terms of the fields (typically just
sums of single particle operators). The other generators, which include
the ``Hamiltonians'' ($P_0$, which is conjugate
to $x^0$ in ET and $P_+$, which is conjugate to the LF-time
$x^+$ in LF quantization) contain interactions among the 
fields and are typically very complicated. 
Generators which leave the initial surface invariant are also 
called {\it kinematic} generators, while the others are called
{\it dynamic} generators. Obviously it is advantageous to have as 
many of the ten generators kinematic as possible. There are
seven kinematic generators on the LF but only six in ET quantization.

The fact that $P_-$, the generator of $x^-$ translations, is
kinematic (obviously it leaves $x^+=0$ invariant!) 
and positive has striking
consequences for the LF vacuum\cite{kent}. For free fields $p^2=m^2$ implies
for the LF energy $p_+ = \left(m^2 + {\vec p}_\perp \right)/2p_-$.
Hence positive energy excitations have positive $p_-$. After the
usual reinterpretation of the negative energy states this implies 
that $p_-$ for a single particle is positive (which makes sense, 
considering that $p_- =\left(p_0-p_3\right)/\sqrt{2}$).
$P_-$ being kinematic
means that it is given by the sum of single particle $p_-$.
Combined with the positivity of $p_-$ this implies that the
Fock vacuum (no particle excitations) is the unique state
with $P_-=0$. All other states have positive $P_-$.
Hence, even in the presence of interactions,
the LF Fock vacuum does not mix with any other state and is
therefore an exact eigenstate of the LF Hamiltonian $P_+$
(which commutes with $P_-$). If one further assumes parity
invariance of the ground state this implies that the Fock
vacuum must be the exact ground state of the fully interacting LF quantum field theory.
\footnote{Practical calculations show that typical LF 
Hamiltonians are either unbounded from below 
or their ground state is indeed the Fock vacuum.}
In sharp contrast to other
formulations of field theory, the LF-vacuum is trivial!
This implies a tremendous technical advantage but also raises
the question whether non-perturbative LF-field theory is
equivalent to conventional field theory, where non-perturbative
effects usually result in a highly nontrivial vacuum structure.
This very deep issue will be the main topic of these lecture
notes.

\section{A First Look at the Light-Front Vacuum}
In the Fock space expansion one starts from the vacuum as the ground
state and constructs physical hadrons by successive application
of creation operators.
In an interacting theory the vacuum is in general an extremely
complicated state and not known a priori. Thus, in general,
a Fock space expansion is not practical because one does not
know the physical vacuum (i.e. the ground state of the
Hamiltonian). In normal coordinates, particularly
in the Hamiltonian formulation, this is a serious obstacle
for numerical calculations.
As is illustrated in Table \ref{tab:vac}, the LF formulation
provides a dramatic simplification at this point.
\begin{table}
\begin{center}
\begin{tabular*}{11.5cm}[t]{@{\extracolsep{\fill}}c|c}
normal coordinates & light-front \\
\rule{6cm}{0.cm}&\rule{6cm}{0.cm}\\
\hline
\multicolumn{2}{c}{free theory}\\[1.5ex]
\multicolumn{2}{c}{
\setlength{\unitlength}{0.6mm}
\special{em:linewidth 0.4pt}
\def\empoint##1{\special{em:point ##1}}
\def\emline##1##2{\special{em:line ##1,##2}}
\begin{picture}(160,65)
\put( 0, 0){\begin{picture}(80,60)( 0, 0)
\put(5,0){\line(1,0){50}}
\put(30,-3){\line(0,1){50}}
\put(60,0){\makebox(0,0){$P_z$}}
\put(75,0){\line(0,1){60}}
\put(30,55){\makebox(0,0){$P^0 = \sqrt{m^2+{\vec{P}}^2}$}}
\put( 5, 27){\circle*{.1}}
\put( 5,27){\circle*{.1}}
  \put(10,22){\circle*{.1}}
  \put(15,18){\circle*{.1}}
  \put(20,14){\circle*{.1}}
  \put(25,11){\circle*{.1}}
  \put(30,10){\circle*{.1}}
  \put(35,11){\circle*{.1}}
  \put(40,14){\circle*{.1}}
  \put(45,18){\circle*{.1}}
  \put(50,22){\circle*{.1}}
  \put(55,27){\circle*{.1}}
\end{picture}}
\put(80, 0){\begin{picture}(80,60)
\put(5,0){\line(1,0){50}}
\put(10,-3){\line(0,1){50}}
\put(60,0){\makebox(0,0){$P_-$}}
\put(10,55){\makebox(0,0)[l]{$P_+ = 
\frac{\T m^2+{\vec P}_{\perp}^2}{\T 2 P_-}$}}
\put(15,49){\circle*{.1}}
\put(20,25){\circle*{.1}}
\put(25,17){\circle*{.1}}
\put(30,13){\circle*{.1}}
\put(35,10){\circle*{.1}}
\put(40,8){\circle*{.1}}
\put(45,7){\circle*{.1}}
\put(50,6){\circle*{.1}}
\put(55,6){\circle*{.1}}
\end{picture}}
\end{picture}
}\\[1.cm]
$ P^0 = {\D\sum\limits_{\vec{k}}} a^\dagger_{\vec{k}} a_{\vec{k}} \sqrt{m^2 + \vec{k}^2 } $ &
$ P_+ = {\D\sum\limits_{k_-,{\vec k}_{\perp}}} 
a^\dagger_{k_-\!,{\vec k}_{\perp}} a_{k_-\!,{\vec k}_{\perp}}
 \frac{ m^2 + {\vec k}_{\perp}^2 }{ 2 k_-} $ \\[1.5ex]
\multicolumn{2}{c}{vacuum (free theory)}\\[1.5ex]
$\D a_{\vec{k}}|0\rangle = 0 $ & $\D a_{k_-\!,k_\perp}|0
\rangle = 0 $\\[1.5ex]
\multicolumn{2}{c}{vacuum (interacting theory)}\\[1.5ex]
many states with $\vec{P}=0$ & $k_- \ge 0$ \\
(e.\,g.\ $a_{\vec{k}}^\dagger
 a_{-\vec{k}}^\dagger|0\rangle$) &
$\hookrightarrow$ only pure zero-mode \\
 & excitations have $P_-=0$\\[1.5ex]
$\hookrightarrow$ $|\tilde{0}\!>$ very complex &
$\hookrightarrow$ $|\tilde{0}\!>$ can only contain \\
 & zero-mode excitations
\end{tabular*}
\end{center}
\caption{Zero Modes and the Vacuum}
\label{tab:vac}
\end{table}
While all components of the momentum in normal coordinates can
be positive as well as negative, the longitudinal LF momentum
$P_-$ is always positive. In free field theory (in normal
coordinates as well as on the LF) the vacuum is the state which
is annihilated by all annihilation operators $a_k$.
In general, in an interacting theory, excited states (excited with
respect to the free Hamiltonian) mix with the trivial vacuum 
(i.e. the free field theory vacuum) state
resulting in a complicated physical vacuum.
Of course, there are certain selection rules and only states with
the same quantum numbers as the trivial vacuum can mix with
this state; for example, states with the same momentum as
the free vacuum (${\vec P}=0$ in normal coordinates,
$P_-=0$, ${\vec P}_\perp =0$ on the LF).
In normal coordinates this has no deep consequences because there
are many excited states which have zero momentum. On the LF
the situation is completely different. Except for pure zero-mode
excitations, i.e. states where only the zero-mode
(the mode with $k_-=0$) is excited, all excited states have
positive longitudinal momentum $P_-$. Thus only these pure zero-mode
excitations can mix with the trivial LF vacuum.
Thus with the exception of the zero-modes the physical LF vacuum
(i.e. the ground state) of an interacting
field theory must be trivial (the only exceptions are pathological
cases, where the LF Hamiltonian is unbounded from below).

Of course, this cannot mean that the vacuum is entirely
trivial. Otherwise it seems impossible to describe
many interesting problems which are related to spontaneous
symmetry breaking within the LF formalism. For example one knows
that chiral symmetry is spontaneously broken in QCD
and that this is responsible for the relatively small mass of
the pions --- which play an important role in strong interaction
phenomena at low energies. What it means is that one has
reduced the problem of finding the LF vacuum to the problem
of understanding the dynamics of these zero-modes.

First this sounds just like merely shifting the problem
about the structure of the vacuum from nonzero-modes
to zero-modes. However, as the
free dispersion relation on the LF, 
\begin{equation}
k_+=\frac{m^2+{\vec k}^2_{\perp}}{2k_-},
\end{equation}
indicates, zero-modes are high energy modes!
Hence it should, at least in principle, be possible
to eliminate these zero-modes systematically
giving rise to an effective LF field theory
\cite{le:ap}. 

\section{Light-Front Hamiltonians without Zero-Modes}
In the following Sections I will discuss several models that were
solved using LF quantization. The solutions to these models
have been obtained by unashamedly omitting explicit zero-mode
degrees of freedom. Nevertheless, physical spectra and condensates
(obtained using current algebra techniques) agree with results obtained
using conventional (non-LF) frameworks.
The models are ordered with increasing complexity.

\subsection{The 't Hooft Model}
From the discussion in the previous section, it first seems that
LF Hamiltonians that do not include zero-mode degrees of
freedom simply give wrong results.
A first indication that this is not necessarily the case was found 
in the
't Hooft model: $QCD_{1+1}(N_c\rightarrow \infty)$ \cite{th:qcd}.
Despite being $1+1$ dimensional, this model has a nonzero fermion
condensate in the chiral limit, since $N_c\rightarrow \infty$.
It is thus interesting to ask what LF quantization predicts
for this case.

The original solution presented by 't Hooft \cite{th:qcd} was obtained
in the LF formulation (quantization \& gauge) did not involve any
zero modes. Furthermore, only a simple principal value prescription
was used to regulate the $q_-=0$ singularity in the gluon
propagator. Nevertheless, the spectrum obtained by 't Hooft agreed
with the spectrum that was obtained later in an ET approach
\cite{qcd2:et}. This is even more surprising if one considers that
the vacuum state in the ET calculation had to be determined by
solving coupled, nonlinear integral equations, whereas the LF vacuum
is just empty space. Already at this point it was clear that the
LF calculation cannot be complete nonsense.

A direct evaluation of the quark condensate in the ET case, gave
a nonzero result in the chiral limit. Since the LF vacuum is the
Fock vacuum, a direct evaluation gave of course zero on the LF.
However, application of current algebra techniques to meson masses 
and coupling constants determined by solving the (zero-mode free)
LF equations gives a nonzero result for the condensate
--- even in the zero quark mass limit:
\begin{eqnarray}
0 &=& \lim_{q \rightarrow 0} iq^\mu \int d^2x e^{iqx}
\langle 0|T\left[\bar{\psi}\gamma_\mu \gamma_5\psi (x)
\bar{\psi}i\gamma_5\psi (0)\right] |0\rangle
\nonumber\\
&=&-\langle 0|\bar{\psi}\psi |0\rangle
-2m_q \int d^2x 
\langle 0|T\left[\bar{\psi}i \gamma_5\psi (x)
\bar{\psi}i\gamma_5\psi (0)\right] |0\rangle .
\nonumber\\
\label{eq:cural}
\end{eqnarray}
Upon inserting a complete set of meson states\footnote{
Because we are working at leading order in $1/N_C$, the
sum over one meson states saturates the operator product
in Eq.(\ref{eq:cural}).}
one thus obtains
\begin{equation}
\langle 0|\bar{\psi}\psi |0\rangle
= -m_q \sum_n \frac{f_P^2(n)}{M_n^2},
\label{eq:naivesum}
\end{equation}
where 
\begin{equation}
f_P(n)\equiv \langle 0|\bar{\psi}i\gamma_5\psi |n\rangle
=\sqrt{\frac{N_C}{\pi}} \frac{m_q}{2}\int_0^1 dx \frac{1}{x(1-x)} \phi_n(x)\ \ 
\end{equation} 
and the wave functions $\phi_n$ and invariant masses $M_n^2$ are
obtained from solving
't Hooft's bound state equation for mesons in 
$QCD_{1+1}$
\begin{equation}
M_n^2 \phi_n(x) = \frac{m_q^2}{x(1-x)}\phi_n(x) + G^2 \int_0^1 dy
\frac{\phi_n(x)-\phi_n(y)}{(x-y)^2}.
\label{eq:thooft}
\end{equation}
The result for $\langle 0|\bar{\psi}\psi |0\rangle$
obtained this way agrees with the ET calculation \cite{qcd2:lfvac}
\footnote{It should be emphasized that the LF calculation preceded
the ET calculation.}.
\begin{figure}
\unitlength1.cm
\begin{picture}(10,6.)(2.9,14.1)
\includegraphics{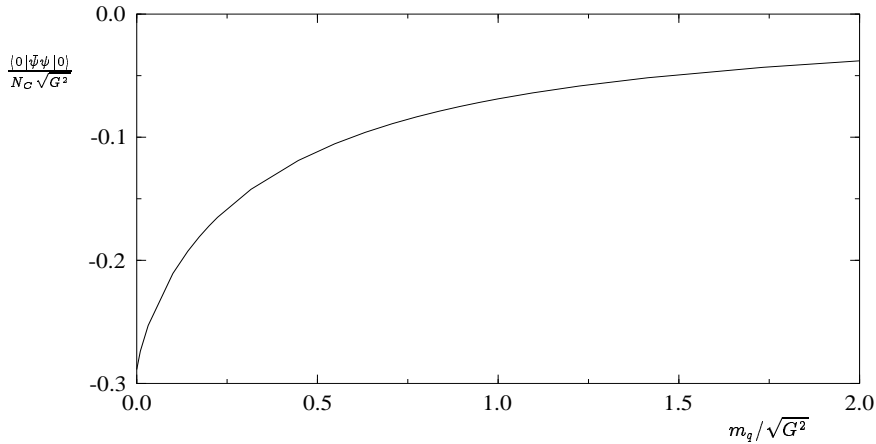}
\end{picture}
\caption{Chiral condensate obtained by evaluating Eq. (\protect\ref{eq:naivesum})
as a function of the quark mass. For nonzero quark mass, the (infinite)
free part has been subtracted. The result agrees for all quark mass with
the calculation done using equal time quantization.
}
\label{fig:thooft}
\end{figure}

This seemingly paradoxical result (peaceful coexistence of a Fock
vacuum and a nonzero fermion condensate) can be understood by 
defining LF quantization through a limiting procedure \cite{le:ap},
where the quantization surface is kept space-like, but being
carefully ``rotated'' to the LF \footnote{Another way of thinking
about this procedure is to imagine a gradual boost to infinite
momentum.}. 
Not all physical quantities behave continuously under this procedure
as the LF is approached. For example, the chiral condensate
$\langle 0|\bar{\psi}\psi|0\rangle $ has a discontinuous LF limit.
On the other hand, the equation of motion for mesons in $QCD_{1+1}$
does have a smooth LF limit. This result explains why the current
algebra relation
gives the right result for the condensate, even though 
$\langle 0|\bar{\psi}\psi|0\rangle $ vanishes when evaluated
directly on the LF: Since the bound state equation for mesons
has a smooth LF limit, both meson masses and coupling constant
can be evaluated directly on the LF. Since the current algebra
relation (\ref{eq:cural}) is a frame independent relation, it can
then be used to extract the condensate from the LF calculation. 
However, since $\langle 0|\bar{\psi}\psi|0\rangle $ has a 
discontinuous LF limit, it would be misleading to draw conclusions
about the vacuum structure from its value obtained directly on the LF.

Another lesson that one should learn from this exercise is that one
should always be careful when defining operators and observables
on the LF. Unless proven otherwise, one should always be prepared that
nontrivial renormalizations occur and that the canonical expressions
are no longer valid.

Finally, a warning should be issued at this point: in the 't Hooft model,
one obtains the correct spectra and (after some detours) even the correct
chiral condensate ``for free'', i.e. by using the naive (canonical)
LF Hamiltonian. It turns out that this is an exception rather than the rule,
i.e. in most theories the canonical LF Hamiltonian yields incorrect results.
This point will be explored in the following sections.

\subsection{Self-Interacting Scalar Fields}
\begin{figure}
\unitlength1.cm
\begin{picture}(10,5.)(-5,0)
\includegraphics{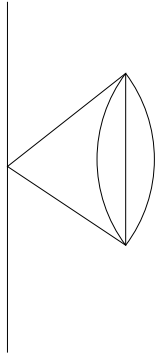}
\end{picture}
\caption{Generalized tadpole (Feynman-) diagram in $\phi^4$ theory.
}
\label{fig:phi4}
\end{figure}

\begin{figure}
\unitlength1.cm
\begin{picture}(10,5.)(-4,-4.5)
\put(-3,-4){\vector(0,1){3.}}
\put(-2.5,-2.5){\makebox(0,0){$x^+$}}
\includegraphics{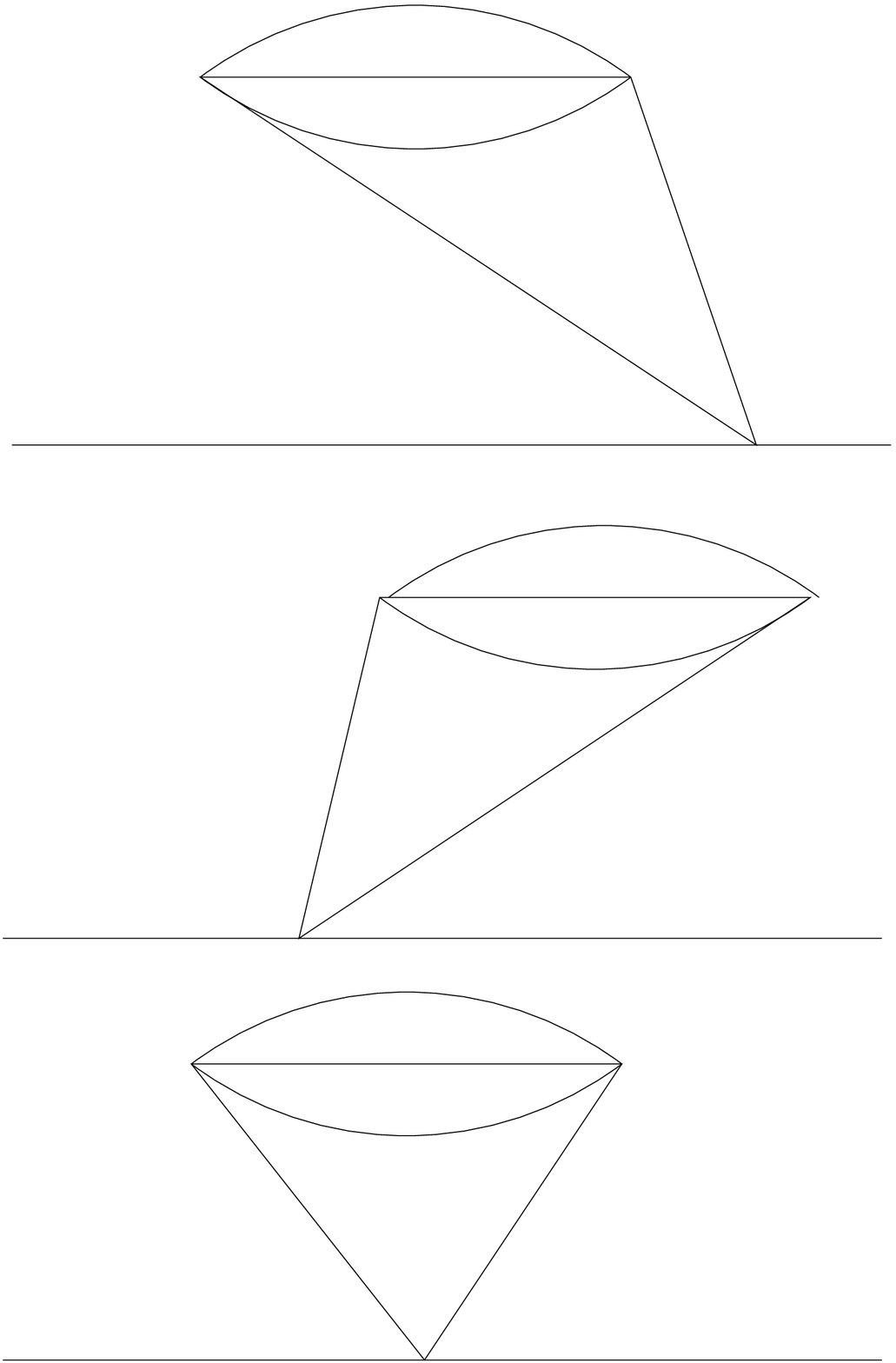}
\end{picture}
\caption{Same as Fig. \protect\ref{fig:phi4} but as LF-time ordered diagrams. At least one of
the vertices has all lines popping out of or disappearing into the vacuum.
}
\label{fig:phi4t}
\end{figure}

In the 't Hooft model discussed above, the naive (canonical)
LF calculation automatically gave the correct meson spectrum.
It turns out that this is an exception rather than the rule!
In most theories one obtains different spectra when one diagonalizes
the canonical ET Hamiltonian and canonical LF Hamiltonian. 
In fact, one does not have to look
very hard to find such examples --- disagreement between LF and ET
calculations arise already with self-interacting scalar fields,
described by the generic Lagrangian 
\begin{equation}
{\cal L}=\frac{1}{2}\partial_\mu \phi \partial^\mu \phi
- V(\phi),
\end{equation}
where
\begin{equation} 
V(\phi)=\sum_{n>1}\frac{\lambda_n}{n!}\phi^n .
\label{eq:vofphi}
\end{equation}
The main difference between the LF formulation and the ET formulation
is that generalized tadpoles (a typical example is shown in Fig.\ref{fig:phi4}), 
i.e. Feynman diagrams where one piece
of the diagram is connected to the rest of the diagram only at one
point, cannot be generated by a LF Hamiltonian: in time ordered
perturbation theory, at least one of the
vertices in a generalized tadpole diagram has all lines coming out
of or disappearing into the vacuum (Fig.\ref{fig:phi4t}) --- 
which is forbidden on the LF
(without zero-modes). 

So the bad news is that all generalized tadpole diagrams are zero
on the LF and they are nonzero in ET quantization, i.e. there is
a difference between the perturbation series generated by the two 
formulations \cite{paul:sg,mb:sg}.

However, there are two good news. The first good news is that in
self-interacting scalar theories, it is only in generalized tadpole
diagrams where such a difference occurs. \footnote{To my knowledge,
there is no strict proof of this result, but it is based on
handwaving arguments as well as on a thorough three loop analysis.} 
The second good news is that all tadpole sub-diagrams are just
constants, i.e. they do not depend on the momenta of any external legs.
The reason is that there is no momentum flowing through them.
Since generalized tadpoles are only constants, their absence can
be compensated for by local counter-terms in the interaction.

In other words, the difference between LF quantization and ET
quantization arises only if one compares calculations done with
the same bare parameters! Suppose one would start with parameters
that have been chosen so that the bare parameters on the LF
include already the counter-terms that are necessary to compensate
for the absence of tadpoles. Then ET and LF formulation should
give the same results for all n-point Green's functions, i.e.
physical observables should be the same. But how can one find
the appropriate counter-terms without having to refer to an ET
calculation? There is nothing easier than that: simply by using
only physical input parameters to fix the bare parameters!
For example, if one matches the physical masses of a few particles
between an ET calculation and a LF calculation then the bare parameters
that one needs in order to get these masses will be different in ET
and LF quantization. The difference will be just such that it compensates
for the absence of tadpoles on the LF and hence all further 
observables will be the same. In other words there is no problem
at all with the LF formulation.

Beyond this happy ending, there is one more very interesting aspect
to this story, which has to do with vacuum condensates. For example,
every generalized tadpole diagram in $\phi^4$ theory is numerically
equal to a diagram that contributes to $\langle 0|\phi^2|0\rangle$.
For example, the tadpole in Fig. \ref{fig:phi4} is proportional to
a term that contributes to $\langle 0|\phi^2|0\rangle$ to second order in $\lambda$.
In fact, after working out the details one finds that the
additional LF counter-term, necessary to obtain equivalence between
ET and LF quantization is a mass counter-term \cite{mb:sg}
\begin{equation}
\Delta m^2 =\lambda \langle 0|\frac{:\phi^2:}{2}|0\rangle ,
\end{equation}
where $\lambda$ is the four point coupling and the vacuum expectation
value (VEV) on the r.h.s. is to be evaluated in normal coordinates.

This result can be readily generalized to an arbitrary polynomial
interaction. One finds the following dictionary: perturbation theory
based on a canonical equal time Hamiltonian with
\begin{equation}
{\cal L}_{int}^{ET}=\sum_n \lambda_n \frac{:\phi^n:}{n!} 
\end{equation} 
and perturbation theory based on a canonical light-front
Hamiltonian with
\begin{equation}
{\cal L}_{int}^{LF}=\sum_n \tilde{\lambda}_n \frac{:\phi^n:}{n!} 
\end{equation} 
are equivalent if 
\begin{equation}
\tilde{\lambda}_n = \sum_{k\leq n} \lambda_{n-k} 
\langle 0|\frac{:\phi^k:}{k!}|0\rangle  .
\label{eq:dict}
\end{equation}
\begin{figure}
\unitlength1.cm
\begin{picture}(10,4.)(1,-7.5)
\includegraphics{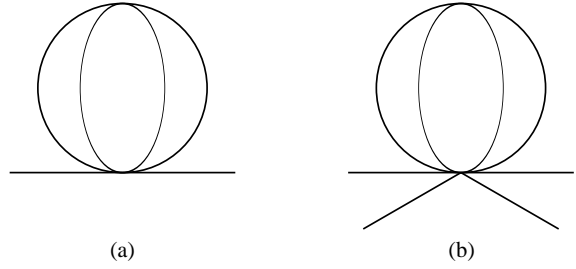}
\end{picture}
\caption{Generalized tadpole diagrams for scalar field theories with
higher polynomial interactions. Both are set to zero in LF quantization without
zero-modes. Both are proportional to $\langle 0|\phi^4|0\rangle$.
The diagram in a.) gives rise to a mass renormalization counter-term and b.)
renormalizes the four-point interaction.
}
\label{fig:basket}
\end{figure}
In Ref.\cite{mb:sg} this fundamental result was derived perturbatively
and the prescription for constructing the effective Hamiltonian was
tested non-perturbatively by calculating physical masses of
of ``mesons'' and solitons in the sine-Gordon model.

At this point it is very tempting to conjecture that this dictionary
(\ref{eq:dict}) also holds for
non-perturbative condensates (such as condensates which arise after
spontaneous symmetry breaking). While a general proof is still
missing, it has indeed been possible to demonstrate for a few
specific models that the conjecture is correct \cite{naus:sg}.

It should also be emphasized that these equivalence considerations
hold irrespective of the number of space-time dimensions, i.e. they
apply to 1+1 as well as 2+1 and 3+1 dimensional theories. One
must only be careful to use commensurate cutoffs when comparing
ET and LF quantized theories. An example would be a transverse 
lattice cutoff, which can be employed both in ET quantization as well
as in LF quantization.

What makes all these results particularly interesting is that they show
how non-perturbative effects can
``sneak'' into the LF formalism and how one can resolve the apparent
conflict between trivial LF vacua and nontrivial vacuum effects.

\subsection{Fermions with Yukawa Interactions}
\begin{figure}
\unitlength1.cm
\begin{picture}(10,4.)(1,5)
\includegraphics{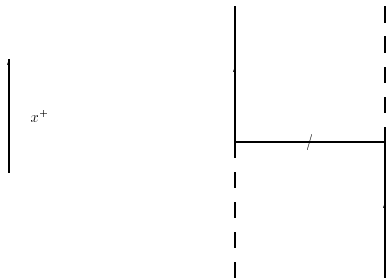}
\end{picture}
\caption{Four point interaction in Yukawa theory that arises after
eliminating $\psi_{(-)}$. The crossed out full line represents the
instantaneous fermion exchange interaction that results from this
elimination. Contracting the boson lines (dashed) yields
a diagram analogous to the tadpole diagrams for self-interacting scalar
fields. 
}
\label{fig:inst}
\end{figure}

Eventually, we are interested to understand chiral symmetry breaking in QCD, 
i.e. we need to understand fermions. As a first step in this direction
let us consider a Yukawa model in 1+1 dimensions
\begin{equation}
{\cal L}=\bar{\psi}\left( i\not \!\!\partial 
-m_F-g\phi \gamma_5 \right)\psi -\phi\left( \Box +m_B^2\right)\phi .
\end{equation}
The main difference between scalar and Dirac fields in the LF formulation is
that not all components of the Dirac field are dynamical: multiplying the
Dirac equation
\begin{equation}
\left( i\not \!\!\partial 
-m_F-g\phi \gamma_5\right)\psi =0
\end{equation}
by $\frac{1}{2}\gamma^+$ yields a constraint equation (i.e. an
``equation of motion'' without a time derivative)
\begin{equation}
i\partial_-\psi_{(-)}=\left(m_F+g\phi\gamma_5\right)\gamma^+\psi_{(+)}
,
\label{eq:constr}
\end{equation}
where
\begin{equation}
\psi_{\pm}\equiv \frac{1}{2}\gamma^\mp \gamma^\pm \psi .
\end{equation}
For the quantization procedure, it is convenient to eliminate
$\psi_{(-)}$ from the classical Lagrangian before imposing
quantization conditions, yielding
\begin{eqnarray}
{\cal L}&=&\sqrt{2}\psi_{(+)}^\dagger \partial_+ \psi_{(+)}
-\phi\left( \Box +m_B^2\right)\phi
-\psi^\dagger_{(+)}\frac{m_F^2}{\sqrt{2}i\partial_-}
\psi_{(+)}
\label{eq:lelim}
\\
&-&\psi^\dagger_{(+)}\left(
g\phi
\frac{m_F\gamma_5}{\sqrt{2}i\partial_-}
+\frac{m_F\gamma_5}{\sqrt{2}i\partial_-}g\phi\right)
\psi_{(+)}
-\psi^\dagger_{(+)}g\phi\frac{1}{\sqrt{2}i\partial_-}
g\phi\psi_{(+)} .
\nonumber
\end{eqnarray}
The rest of the quantization procedure very much resembles the procedure
for self-interacting scalar fields.
In particular, we must be careful about generalized tadpoles,
which might cause additional counter-terms in the LF Hamiltonian.
In the Yukawa model one usually (i.e. in a covariant formulation)
does not think about tadpoles. However, after eliminating $\psi_{(-)}$,
we are left with a four-point interaction in the Lagrangian, which does
give rise to time-ordered diagrams that resemble tadpole diagrams 
(Fig.\ref{fig:inst}).
In fact, the four-point interaction gives rise to diagrams where
a fermion emits a boson, which may or may not self-interact, and then
re-absorb the boson at the same LF-time. \footnote{There are also tadpoles,
where the fermions get contracted. But those only give rise to an additional
boson mass counter-term, but not to a non-covariant counter-term that we
investigate here.}
As we discussed in detail in the previous section, such interactions
cannot be generated by a LF Hamiltonian, i.e. the LF formalism
defines such tadpoles to be zero.

For self-interacting scalar fields, the difference between ET and LF
perturbation theory which thus results can be compensated by a
redefinition of parameters that appear already in the Lagrangian.
In the Yukawa model, the situation is a little more complicated.
The missing tadpoles have the same operator/Lorentz structure as the
so called kinetic mass term
\begin{equation}
{\cal P}^-_{kin} =\psi^\dagger_{(+)}\frac{m_F^2}{\sqrt{2}i\partial_-}
\psi_{(+)} .
\label{eq:pkin}
\end{equation}
One obtains this result by contracting the two scalar fields in the
four-point interaction. More details can be found in Ref. \cite{mb:alex}.
The important point here is that there is no similar counter-term
for the term linear in the fermion mass $m_F$. Thus the difference
between ET and LF quantization cannot be compensated by tuning the
bare masses differently. The correct procedure requires to renormalize
the kinetic mass term (the term $\propto m_F^2$) and the vertex mass term
(the term $\propto m$) independent from each other. More explicitly this
means that one should make an ansatz for the renormalized LF Hamiltonian
density of the form
\begin{eqnarray}
{\cal P}^-&=&
\frac{m_B^2}{2}\phi^2
+\psi^\dagger_{(+)}\frac{c_2}{\sqrt{2}i\partial_-}
\psi_{(+)}
+c_3\psi^\dagger_{(+)}\left(
\phi
\frac{\gamma_5}{\sqrt{2}i\partial_-}
+\frac{\gamma_5}{\sqrt{2}i\partial_-}g\phi\right)
\psi_{(+)} \nonumber\\
&+&c_4\psi^\dagger_{(+)}\phi\frac{1}{\sqrt{2}i\partial_-}
\phi\psi_{(+)} ,
\label{eq:pren}
\end{eqnarray}
where the $c_i$ do not necessarily satisfy the canonical relation
$c_3^2=c_2c_4$. However, this does not mean that the $c_i$ are completely
independent from each other. In fact, only for specific combinations of
$c_i$ will Eq.(\ref{eq:pren}) describe the Yukawa model. It is only that
we do not know the relation between the $c_i$.

Thus the bad new is that the number of parameters in the LF Hamiltonian
has increased by one (compared to the Lagrangian). The good news is that
a wrong combination of $c_i$ will in general give rise to a parity 
violating theory. \footnote{As an example, consider the free theory,
where the correct relation ($c_3^2=c_2c_4$) follows from a covariant
Lagrangian. Any deviation from this relation can be described on
the level of the Lagrangian (for free massive fields, equivalence
between LF and covariant formulation is not an issue)
by addition of a term of the form $\delta {\cal L} =
\bar{\psi}\frac{\gamma^+}{i\partial^+}\psi$, which is obviously 
parity violating, since parity transformations result in
$A^\pm \rightarrow A^\mp$ for Lorentz vectors $A^\mu$; i.e. $\delta {\cal L}
\rightarrow \bar{\psi}\frac{\gamma^-}{i\partial^-}\psi \neq \delta {\cal L}$.}
This is good news because one can thus use parity
invariance for physical observables as an additional renormalization
condition to determine the additional ``free'' parameter.

In fact, in Ref. \cite{mb:parity}, it was shown that 
utilizing parity constraints as renormalization conditions
is practical. The observable considered in that work was the
vector transition form factor (in a scalar Yukawa theory in 1+1 dimensions) 
between physical meson states of opposite C-parity (and thus
supposedly opposite parity)

\begin{equation}
\langle p^\prime, n| j^\mu | p, m \rangle
\stackrel{!}{=} \varepsilon^{\mu \nu}q_\nu F_{mn}(q^2),
\label{eq:form}
\end{equation}
where $q=p^\prime -p$. When writing the r.h.s. in terms of
one invariant form factor, use was made of both vector current
conservation and parity invariance. A term proportional
to $p^\mu + {p^\prime}^\mu$ would also satisfy current conservation,
but has the wrong parity. A term proportional to 
$\varepsilon^{\mu \nu}p_\nu + p_\nu^\prime$ has the right parity,
but is not conserved and a term proportional to $q^\mu$ is both
not conserved and violates parity. Other vectors do not exist for
this example.
The Lorentz structure in Eq. (\ref{eq:form}) has nontrivial
implications even if we consider only the ``good'' component
of the vector current \footnote{In the context of LF calculations,
currents that are bilinear in the dynamical component $\psi_{(+)}$
are usually easiest to renormalize and calculate. Other combinations,
such as $\psi_{(-)}^\dagger \psi_{(-)}$ involve interactions when
expressed in terms of the dynamical components and are thus terrible
to handle --- hence they are often referred to as ``bad'' components.},
yielding
\begin{equation}
\frac{1}{q^+}\langle p^\prime, n| j^+ | p, m \rangle
=F_{mn}(q^2).
\label{eq:formplus}
\end{equation}
That this equation implies nontrivial constraints can be seen
as follows: as a function of the longitudinal momentum transfer
fraction $x\equiv q^+/p^+$, the invariant momentum transfer 
reads ($M_m^2$ and $M_n^2$ are the invariant masses of the
in and outgoing meson)
\begin{equation}
q^2=x\left(M_m^2 -\frac{M_n^2}{1-x}\right)
\end{equation}

\begin{figure}
\unitlength1.cm
\begin{picture}(10,15.7)(1.7,1.9)
\includegraphics{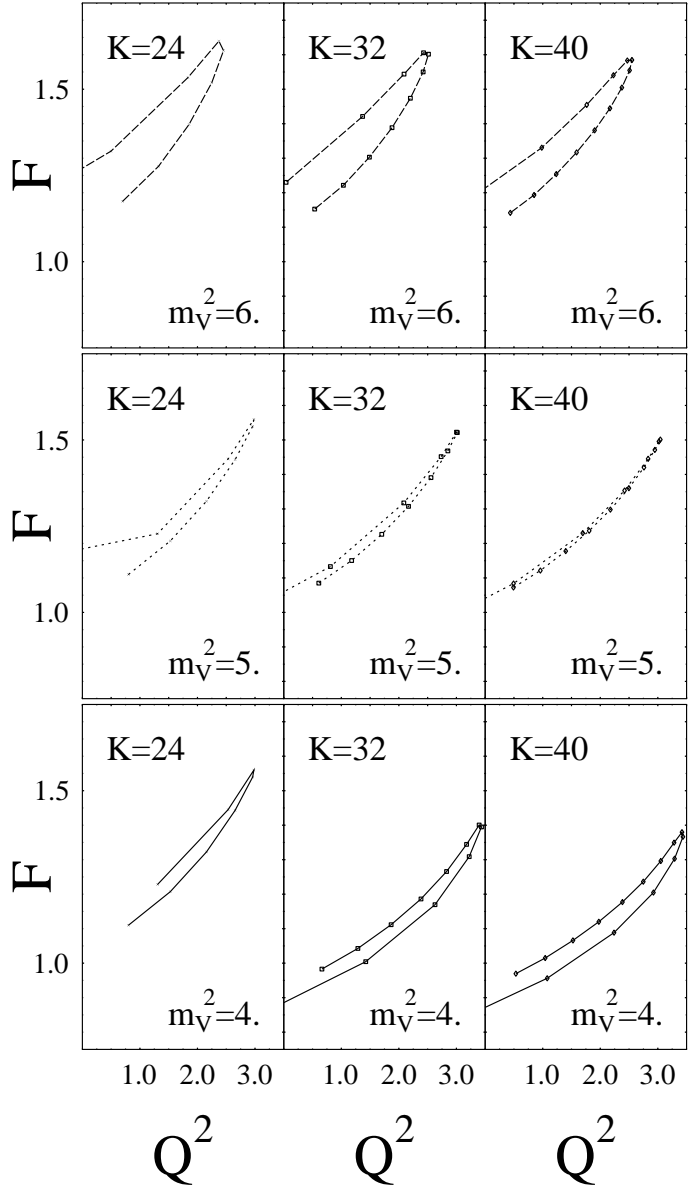}
\end{picture}
\caption{
Inelastic transition form factor (\protect\ref{eq:formplus}) between the
two lightest meson states of the Yukawa model, calculated
for various vertex masses $m_v$ and for various DLCQ parameters
$K$. The physical masses for the fermion and the scalar meson
have been renormalized to the values 
$\left(m_F^{phys}\right)^2=\left(m_F^{phys}\right)^2=4$.
All masses and momenta are in units of
$\protect\sqrt{\lambda} = \protect\sqrt{c_4/2\pi}$. In this example,
only for $m_V^2\approx 5$ one obtains a form factor that is a unique
function of $Q^2$, i.e. only for $m_V^2\approx 5$, the result is
consistent with Eq. (\protect\ref{eq:formplus}). Therefore, only for
this particular value of the vertex mass, is the matrix element of the current operator consistent with both parity and current conservation.
}
\label{fig:parity}
\end{figure}

Typically, there are two values of $x$ that lead to the same
value of $q^2$. It is highly nontrivial to obtain the same
form factor for both values of $x$. In Ref. \cite{mb:parity},
the coupling as well as the 
physical masses of both the fermion and the lightest boson where
kept fixed, while the ``vertex mass'' was tuned (note that this
required re-adjusting the bare kinetic masses). Figure \ref{fig:parity}
shows a typical example. In that example, the calculation of the
form factor was repeated for three values of the DLCQ parameter K
(24, 32 and 40) in order to make sure that numerical approximations
did not introduce parity violating artifacts.
For the ``magic value''
of the vertex mass one finds that the parity condition 
(\ref{eq:formplus}) is satisfied over the whole range of $q^2$
considered. This provides a strong self-consistency check, since
there is only one free parameter, but the parity condition is not
just one condition but a condition for every single value of
$q^2$ (i.e. an infinite number of conditions). In other words,
keeping the vertex mass independent from the kinetic mass is not
only necessary, but also seems sufficient in order to properly 
renormalize Yukawa$_{1+1}$.

\newpage 
\subsection{A Model with Spontaneous Breakdown of Chiral Symmetry}
The discussion in the previous section showed that the renormalization
of fermions on the LF requires the introduction of non-covariant
counter-terms. The question is: are a finite number of such
counter-terms to the LF Hamiltonian sufficient to describe a physical
situation with spontaneous breakdown of chiral symmetry?
In order to investigate this question, I was looking for a nontrivial
model where the mechanism for chiral symmetry breaking ($\chi$SB)
is understood\footnote{By the way, this excludes $QCD_{3+1}$ since
$\chi$SB is not yet understood there --- even in ``normal'' coordinates.}
and which is also suitable for a LF formulation.
In addition, the model should allow to work with the same cutoffs
and approximations in the conventional formulation
and the LF formulation, since this facilitates a comparison
the two frameworks at each possible step in the calculations.

One model that has all the above features 
consists of fermions coupled to the transverse component of a 
non-Abelian vector field and which is described by
the Lagrangian
\begin{equation}
{\cal L}=\bar{\psi}_k\left( i\partial \!\!\!\!\!\!\not\,\,\, -m\right)\psi_k
-\frac{g}{\sqrt{N_C}}\bar{\psi}_k {\vec \gamma}_\perp {\vec A}_\perp^{kl}
\psi_l - \frac{1}{2}{\vec A}_\perp^{kl}\left( 
\Box + \lambda^2\right){\vec A}_\perp^{kl} .
\label{eq:lbizarre}
\end{equation}
$k,l$ can be interpreted as ``color'' indices and the $N\rightarrow
\infty$ limit is taken.
Without going into details, the motivation for choosing such a bizarre 
model was the following: the coupling of the fermion to a vector
field is chirally invariant; the longitudinal components of the vector
field was omitted completely since vector couplings with $A^+\neq 0$
are notoriously difficult on the LF; a non-Abelian  not-self-interacting
was chosen since this results in a solvable model in the
$N_C\rightarrow \infty$ limit.

Despite certain similarities, the above model is not a gauge theory
and there is nothing wrong with imposing momentum cutoffs.
A cutoff that can be used both in ``normal coordinates'' as
well as on the LF is a transverse momentum cutoff
(sharp momentum cutoff or transverse lattice or similar). 

In normal coordinates, one can easily solve above model by
solving the Dyson-Schwinger (DS) equations in the rainbow approximation (exact in the $N_C\rightarrow \infty$ limit).
What we will demonstrate in the following
subsections is that a standard LF calculation \footnote{Such as 
DLCQ without zero-modes, but in the limit of large harmonic 
resolution.} with appropriate counter-terms yields the
same spectrum as a conventional calculation. 
This result holds for all values of the quark mass \footnote{Except strictly zero, but this
point can easily be approached via a limiting procedure.}.
which makes it particularly interesting since the model
that we are considering exhibits spontaneous breakdown of chiral symmetry in the limit $m\rightarrow 0$.

Since the details of this proof are rather lengthy and formal,
readers not interested in details should immediately proceed
to the Summary.

\subsubsection{Schwinger-Dyson Solution:}

Because the above toy model lacks full covariance 
(there is no symmetry relating longitudinal and transverse coordinates)
the full fermion propagator is of the form
\begin{equation}
S_F(p^\mu) = \not \! \! p_L S_L({\vec p}_L^2,{\vec p}_\perp^2)
+ \not \! \! \;p_\perp S_\perp({\vec p}_L^2,{\vec p}_\perp^2)+S_0({\vec p}_L^2,{\vec p}_\perp^2),
\end{equation}
where $\not \! \! k_L \equiv k_0\gamma^0 + k_3 \gamma^3$ and
$\not \! \! k_\perp \equiv k_1\gamma^1 + k_2 \gamma^2$. 
On very general grounds, it should always be possible to write
down a spectral representation for $S_F$\footnote{What we need
is that the Green's functions are analytic except for poles
and that the location of the poles are consistent with
longitudinal boost invariance (which is manifest in our model).
The fact that the model is not invariant under transformations
which mix $p_L$ and $p_\perp$ does not prevent us from writing
down a spectral representation for the dependence on $p_L$.
}
\begin{equation}
S_i({\vec p}_L^2,{\vec p}_\perp^2) = \int_0^\infty dM^2 \frac{\rho_i(M^2,{\vec p}_\perp^2)}
{{\vec p}_L^2-M^2+i\varepsilon},
\label{eq:sansatz}
\end{equation}
where $i=L,\perp,0$.
Note that this spectral representation differs from what one
usually writes down as a spectral representation in that we are not 
assuming full covariance here.
Note that in a covariant theory, one usually writes down spectral
representations in a different form, namely
$S=\int_0^\infty d\tilde{M}^2 \tilde{\rho}(\tilde{M}^2)/({\vec p}_L^2-{\vec p}_\perp^2-\tilde{M}^2)$, i.e.
with ${\vec p}_\perp^2$ in the denominator. This is a special case of
Eq. (\ref{eq:sansatz}) with $\rho(M^2,{\vec p}_\perp^2)=
\int_0^\infty d\tilde{M}^2 \tilde{\rho}
(\tilde{M}^2)\delta(M^2-\tilde{M}^2-{\vec p}_\perp^2)$.

Using ansatz from above (\ref{eq:sansatz})
for the spectral densities, one finds for the 
self-energy
\begin{eqnarray}
\Sigma(p^\mu) &\equiv& ig^2 \int \frac{d^4k}{(2\pi )^4} {\vec \gamma}_\perp
S_F(p^\mu-k^\mu){\vec \gamma}_\perp \frac{1}{k^2-\lambda^2+i\varepsilon}
\nonumber\\
&=& \not \! \! p_L\Sigma_L({\vec p}_L^2,{\vec p}_\perp^2) + 
\Sigma_0({\vec p}_L^2,{\vec p}_\perp^2),
\label{eq:sd1}
\end{eqnarray}
where
\begin{eqnarray}
\Sigma_L({\vec p}_L^2,{\vec p}_\perp^2) &=& g^2 \!\!\int_0^\infty \!\!\!\!\!dM^2 \!\int_0^1\!\!\!dx 
\!\!\!\int \!\frac{d^2k_\perp}{8\pi^3} 
\frac{(1-x)\rho_L(M^2, ({\vec p}-{\vec k})_\perp^2)}{D}
\nonumber\\
\Sigma_0({\vec p}_L^2,{\vec p}_\perp^2) &=& -g^2 \!\!\int_0^\infty \!\!\!\!\!dM^2 \!\int_0^1\!\!\!dx 
\!\int \!\frac{d^2k_\perp}{8\pi^3} 
\frac{\rho_0(M^2, ({\vec p}-{\vec k})_\perp^2)}{D}.
\nonumber\\
\label{eq:sd2}
\end{eqnarray}
and
\begin{equation}
D=x(1-x){\vec p}_L^2 - xM^2
-(1-x)\left({\vec k}_\perp^2+\lambda^2\right)
\end{equation}
Note that $\Sigma_\perp$ vanishes, since $\sum_{i=1,2} \gamma_i \gamma_j
\gamma_i=0$ for $j=1,2$.
Self-consistency then requires that
\begin{equation}
S_F = \frac{1}{\not \! \! p_L\left[1-\Sigma_L({\vec p}_L^2,{\vec p}_\perp^2) \right]
+ \not \! \! p_\perp - \left[m+\Sigma_0({\vec p}_L^2,{\vec p}_\perp^2)\right]}
\label{eq:sd3}
\end{equation}
In the above equations we have been sloppy about cutoffs in order
to keep the equations simple, but this can be easily remedied by
multiplying each integral by a cutoff on the fermion momentum, such as
$\Theta\left(\Lambda^2_\perp-({\vec p}-{\vec k})_\perp^2\right)$
In principle, the set of equations 
[Eqs. (\ref{eq:sd1}),(\ref{eq:sd2}),(\ref{eq:sd3})]
can now be used to determine the spectrum of the model.
But we are not going to do this here since we are more interested
in the LF solution to the model. However, we would still like
to point out that, for large enough $g$, one obtains a self-consistent
numerical solution to the Euclidean version of the model which
has a non-vanishing scalar piece --- even for vanishing current
quark mass $m$, i.e. chiral symmetry is spontaneously
broken and a dynamical mass is generated for the fermion in this model.

\subsubsection{LF Solution:}

A typical framework that people use when solving LF quantized
field theories is discrete light-cone quantization (DLCQ) 
\cite{pa:dlcq}. Since it is hard to take full advantage of the
large $N_C$ limit in DLCQ, we prefer to use a Green's function 
framework based on a 4 component formulation of the model. 
The Green's function approach has the advantage that in the above
toy model more things can be done analytically. In addition, it
allows us to work in the continuum limit, where a comparison with
the Schwinger-Dyson calculation can be done without further extrapolations.
I should emphasize that we did verify that the Green's function
approach (with momentum integrals discretized)
yielded the same physical masses as a DLCQ calculation \cite{mb:hala}.

In a LF formulation of the model, the fermion propagator
(to distinguish the notation from the one above, we denote
the fermion propagator by $G$ in this subsection) should be of the form
\footnote{Note that in a LF formulation, $G_+$ and $G_-$ are
not necessarily the same.}
\begin{eqnarray}
G(p^\mu) &=& \gamma^+ p^-G_+(2p^+p^-,{\vec p}_\perp^2)
+\gamma^- p^+G_-(2p^+p^-,{\vec p}_\perp^2)
\nonumber\\
& &+ \not \!\!k_\perp G_\perp(2p^+p^-,{\vec p}_\perp^2)+
G_0(2p^+p^-,{\vec p}_\perp^2).
\label{eq:lf1}
\end{eqnarray}
Again we can write down spectral representations
\begin{eqnarray}
G_i(2p^+p^-,{\vec p}_\perp^2) = \int_0^\infty dM^2
\frac{\rho_i^{LF}(M^2,{\vec p}_\perp^2)}
{2p^+p^--M^2+i\varepsilon},
\label{eq:speclf}
\end{eqnarray}
where $i=+,-,\perp,0$. This requires some explanation:
On the LF, one might be tempted to allow for two terms
in the spectral decomposition that are proportional to
$\gamma^+$, namely
\begin{equation}
tr(\gamma^-G)\propto
\int_0^\infty dM^2
\frac{p^-\rho_a(M^2,{\vec p}_\perp^2)+\frac{1}{p^+}\rho_b(M^2,{\vec p}_\perp^2)}
{2p^+p^--M^2+i\varepsilon}.
\label{eq:rhoab}
\end{equation}
However, upon writing 
\begin{equation}
\frac{1}{p^+}=\frac{1}{p^+M^2}\left(M^2-2p^+p^-\right)+\frac{2p^-}{M^2}
\end{equation}
one can cast Eq. (\ref{eq:rhoab}) into the form
\begin{eqnarray}
tr(\gamma^-G)&\propto&
\int_0^\infty dM^2p^-
\frac{\rho_a(M^2,{\vec p}_\perp^2)+\frac{2}{M^2}\rho_b(M^2,{\vec p}_\perp^2)
}{2p^+p^--M^2+i\varepsilon}
\nonumber\\
& &-\frac{1}{p^+}\int_0^\infty dM^2
\frac{\rho_b(M^2,{\vec p}_\perp^2)}{M^2},
\label{eq:rhoaab}
\end{eqnarray}
which is of the form in Eq.(\ref{eq:speclf}) plus an energy independent
term. The presence of such an additional energy independent
term would spoil the high energy behavior of the model \cite{brazil}:
In a LF Hamiltonian, not all coupling constants are arbitrary.
In many examples, 3-point couplings and the 4-point couplings
must be related to one another so that the high energy behavior
of scattering via the 4-point interaction and via the iterated
3-point interaction cancel \cite{brazil}. If one does not
guarantee such a cancelation then the high-energy behavior of the
LF formulation differs from the high-energy behavior in covariant
field theory and in addition one often also gets a spectrum that is unbounded
from below. In Eq. (\ref{eq:rhoaab}), the energy independent
constant appears if the coupling constants of the "instantaneous 
fermion exchange" interaction in the LF Hamiltonian and the 
boson-fermion vertex are not properly balanced.
In the following we will assume that one has started with an
ansatz for the LF Hamiltonian with the proper high-energy behavior,
i.e. we will assume that there is no such energy independent 
piece in Eq. (\ref{eq:rhoaab}).

The LF analog of the self-energy equation is obtained by
starting from an expression similar to Eq.(\ref{eq:sd2}) and
integrating over $k^-$. One obtains
\begin{equation}
\Sigma^{LF} = \gamma^+\Sigma_+^{LF}+\gamma^-\Sigma_-^{LF}
+\Sigma_0^{LF},
\end{equation}
where
\begin{eqnarray}
\!\Sigma_+^{LF}\!(p) \!&=&\! g^2 \!\!\!\int_0^\infty \!\!\!\!\!\!dM^2 
\!\!\!\int_0^{p^+}\!\!\!\!\!\!\!dk^+ 
\!\!\!\!\int \!\!\frac{d^2k_\perp}{16\pi^3} 
\frac{\!\!\left(\!p^-\!\!-\frac{\lambda^2+{\vec k}_\perp^2}
{2k^+}\!\right)\!\rho_+^{LF}(M^2\!\!, 
({\vec p}-{\vec k})_\perp^2)}{k^+(p^+-k^+)D^{LF}}
+ CT
\nonumber\\
\!\Sigma_-^{LF}\!(p) \!&=&\! g^2 \!\!\!\int_0^\infty \!\!\!\!\!\!dM^2 \!\!\!\int_0^{p^+}\!\!\!\!\!\!\!dk^+
\!\!\!\!\int \!\!\frac{d^2k_\perp}{16\pi^3} 
\frac{\left(p^+-k^+\right)\rho_-^{LF}(M^2, ({\vec p}-{\vec k})_\perp^2)}{k^+(p^+-k^+)D^{LF}}
\nonumber\\
\!\Sigma_0^{LF}\!(p) \!&=&\! -g^2 \!\!\!\int_0^\infty \!\!\!\!\!\!dM^2 \!\!\!\int_0^{p^+}\!\!\!\!\!\!\!dk^+
\!\!\!\!\int \!\!\frac{d^2k_\perp}{16\pi^3} 
\frac{\rho_0^{LF}(M^2, ({\vec p}-{\vec k})_\perp^2)}{k^+(p^+-k^+)D^{LF}}.
\label{eq:lf2}
\end{eqnarray}
where
\begin{equation}
D^{LF}=p^- - \frac{M^2}{2(p^+-k^+)} - \frac{\lambda^2+{\vec k}_\perp^2}{2k^+}
\end{equation}
and CT is an energy ($p^-$)
independent counter-term. The determination of this counter-term, such
that one obtains a complete equivalence with the Schwinger Dyson 
approach, is in fact the main achievement of this paper.
First we want to make sure that the counter-term renders the self-energy
finite. This can be achieved by performing a ``zero-energy subtraction''
with a free propagator,
analogous to adding self-induced inertias to a LF Hamiltonian, yielding
\begin{equation}
CT= g^2 \int_0^{p^+}\!\!\!\!\!\!\!dk^+ 
\!\!\!\!\int \!\!\frac{d^2k_\perp}{16\pi^3} 
\frac{\!\!\frac{\lambda^2+{\vec k}_\perp^2}{2k^+}\!}{k^+(p^+-k^+)D_0^{LF}}
+\frac{\Delta m^2_{ZM}}{2p^+},
\label{eq:ct1}
\end{equation}
where
\begin{equation}
D_0^{LF}= - \frac{M_0^2+({\vec p}-{\vec k})_\perp^2}{2(p^+-k^+)} - \frac{\lambda^2+{\vec k}_\perp^2}{2k^+}
\end{equation}
and where we denoted the finite piece by $\Delta m^2_{ZM}$ (for {\it zero-mode}), since
we suspect that it arises from the dynamics of the zero-modes.
$M_0^2$ is an arbitrary scale parameter. We will construct
the finite piece ($\Delta m^2_{ZM}$) so that there is no 
dependence on $M_0^2$ left in CT in the end.

At this point, only the infinite part of $CT$ is unique \cite{brazil}, since it
is needed to cancel the infinity in the $k^+$ integral
in Eq. (\ref{eq:lf2}), while the
finite (w.r.t. the $k^+$ integral) piece (i.e. $\Delta m^2_{ZM}$) seems arbitrary. \footnote{Note that what we called the "finite piece"
w.r.t. the $k^+$ integral is still divergent when one integrates over 
$d^2k_\perp$ without a cutoff!}
Below we will show that it is not arbitrary and only
a specific choice for $\Delta m^2_{ZM}$
leads to agreement between the SD and the LF approach.

Note that the equation for the self-energy can also be written in the
form
\begin{eqnarray}
\!\Sigma_+^{LF}\!(p) \!&=&\! g^2  
\!\!\!\int_0^{p^+}\!\!\!\!\frac{dk^+}{k^+} 
\!\!\!\!\int \!\!\frac{d^2k_\perp}{8\pi^3} 
p^-_F
G_+\left(2p^+_Fp^-_F,{\vec p}_{\perp F}^2\right)
+ CT
\nonumber\\
\!\Sigma_-^{LF}\!(p) \!&=&\! g^2  
\!\!\!\int_0^{p^+}\!\!\!\!\frac{dk^+}{k^+}
\!\!\!\!\int \!\!\frac{d^2k_\perp}{8\pi^3} 
p^+_F
G_-\left(2p^+_Fp^-_F,{\vec p}_{\perp F}^2\right)
\nonumber\\
\!\Sigma_0^{LF}\!(p) \!&=&\! -g^2 
\!\!\!\int_0^{p^+}\!\!\!\!\frac{dk^+}{k^+}
\!\!\!\!\int \!\!\frac{d^2k_\perp}{8\pi^3} 
G_0\left(2p^+_Fp^-_F,{\vec p}_{\perp F}^2\right),
\label{eq:lf2b}
\end{eqnarray}
where
\begin{eqnarray}
p^+_F&\equiv& p^+-k^+ \nonumber\\
p^-_F&\equiv& p^--\frac{\lambda^2+{\vec k}_\perp^2}{2k^+}\nonumber\\
{\vec p}_{\perp F} &\equiv&{\vec p}_\perp-{\vec k}_\perp
\end{eqnarray}
One can prove this by simply comparing expressions! 
Bypassing the use of the spectral function greatly simplifies
the numerical determination of the Green's function in a self-consistent
procedure.

\subsubsection{Comparing the LF and SD solutions:}
Motivated by considerations in Ref.\cite{mb:adv}, we make the
following ansatz for ZM:
\begin{equation}
\Delta m^2_{ZM} = g^2\int_0^\infty \!\!\!\!dM^2 \!\!\int 
\!\!\frac{d^2k_\perp}{8\pi^3}
\rho_+^{LF}(M^2,{\vec p}_{F\perp}^2) \ln \frac{M^2}{M_0^2+{\vec p}_{F\perp}^2}.
\label{eq:zm}
\end{equation}
The motivation for this particular ansatz becomes obvious one
we rewrite the expression for $\Sigma_+^{LF}$:
For this purpose, we first note that
\begin{eqnarray}
\frac{p^--\frac{\lambda^2+{\vec k}_\perp^2}{2k^+}}{k^+(p^+-k^+)D^{LF}}
&+&
\frac{\frac{\lambda^2+{\vec k}_\perp^2}{2k^+}}{k^+(p^+-k^+)D_0^{LF}}
\\
= \frac{p^-\frac{p^+-k^+}{p^+}}{k^+(p^+-k^+)D^{LF}}
&-& \frac{1}{p^+}\frac{\partial}{\partial k^+} \ln \left[\frac{D^{LF}}{D_0^{LF}}\right].
\nonumber
\end{eqnarray}
Together with the normalization condition\\ 
$\int_0^\infty dM^2 
\rho_+^{LF}(M^2,{\vec k}_\perp^2)=1$, this implies
\begin{eqnarray}
\!\Sigma_+^{LF}\!(p) \!&=&\! g^2\frac{p^-}{p^+} \!\!\!\int_0^\infty \!\!\!\!\!\!dM^2 
\!\!\!\int_0^{p^+}\!\!\!\!\!\!\!dk^+ 
\!\!\!\!\int \!\!\frac{d^2k_\perp}{16\pi^3} 
\frac{\!\!\left(p^+-k^+\right)\!\rho_+^{LF}(M^2\!\!, ({\vec p}-{\vec k})_\perp^2)}{k^+(p^+-k^+)D^{LF}}
\nonumber\\
& &-\frac{g^2}{2p^+} \int_0^\infty \!\!\!\!\!\!dM^2 \!\!\!\int \!\!\frac{d^2k_\perp}{8\pi^3} 
\rho_+^{LF}(M^2,{\vec p}_{F\perp}^2)\ln \frac{M^2}{M_0^2+{\vec p}_{\perp F}^2 } 
+\frac{\Delta m^2_{ZM}}{2p^+}
\nonumber\\
&=&\! g^2\frac{p^-}{p^+} \!\!\!\int_0^\infty \!\!\!\!\!\!dM^2 
\!\!\!\int_0^{p^+}\!\!\!\!\!\!\!dk^+ 
\!\!\!\!\int \!\!\frac{d^2k_\perp}{16\pi^3} 
\frac{\!\!\left(p^+-k^+\right)\!\rho_+^{LF}(M^2\!\!, ({\vec p}-{\vec k})_\perp^2)}{k^+(p^+-k^+)D^{LF}},
\end{eqnarray}
where we used our particular ansatz for $\Delta m^2_{ZM}$ [Eq. (\ref{eq:zm})].
Thus, with our particular choice for the finite piece of the kinetic
energy counter term, the expressions for $\Sigma_+^{LF}$ and $\Sigma_-^{LF}$
are almost the same --- the only difference being the replacement of
$\rho_+^{LF}$ with $\rho_-^{LF}$ and an overall factor of $p^-/p^+$.
Furthermore, and this is the most important result of this paper, 
a direct comparison (take $x=k^+/p^+$) shows that the same spectral
densities that provide a self-consistent solution to the SD
equations (\ref{eq:sd2}) also yield a self-consistent solution to the
LF equations, provided one chooses
\begin{eqnarray}
\rho_+^{LF}(M^2,{\vec k}_\perp^2) &=&\rho_-^{LF}(M^2,{\vec k}_\perp^2)
=\rho_L(M^2,{\vec k}_\perp^2)\nonumber\\
\rho_0^{LF}(M^2,{\vec k}_\perp^2) &=&\rho_0(M^2,{\vec k}_\perp^2).
\end{eqnarray}
In particular, the physical masses
of all states (in the sector with fermion number one)
must be the same in the SD and the LF framework.

In the formal considerations above, we found it convenient to
express $\Delta m^2_{ZM}$ in terms of the spectral density.
However, this is not really necessary since one can express
it directly in terms of the Green's function
\begin{eqnarray}
\Delta m^2_{ZM}&=&g^2p^+\!\!\!\!\int_{-\infty}^0\!\!\!\!\!\!dp^-\!\!\!\!
\int \!\!\frac{d^2p_\perp}{
4\pi^3} \!\!\left[
G_+(2p^+p^-,{\vec p_\perp}^2)
\label{eq:dmgreen}
-\frac{1}{2p^+p^--{\vec p}_\perp^2-M_0^2}\right] .
\end{eqnarray} 
Analogously, one can also perform a "zero-energy subtraction" in
Eq. (\ref{eq:lf2b}) with the full Green's function, i.e.
by choosing
\begin{equation}
CT=-g^2  
\!\!\!\int_0^{p^+}\!\!\!\!\frac{dk^+}{k^+} 
\!\!\!\!\int \!\!\frac{d^2k_\perp}{8\pi^3} 
\tilde{p}^-_F
G_+\left(2p^+_F\tilde{p}^-_F,{\vec p}_{\perp F}^2\right),
\label{eq:ctilde}
\end{equation}
with $\tilde{p}^-_F=-(\lambda^2+{\vec k}_\perp^2)/2k^+$.
This expression turns out to be very useful when constructing the
self-consistent Green's function solution. 
We used both ans\"atze [Eqs. (\ref{eq:dmgreen}) and 
(\ref{eq:ctilde})] to determine the physical masses of the
dressed fermion. In both cases, numerical agreement with
the solution to the Euclidean SD equations was obtained.

Note that, in a canonical
LF calculation (e.g. using DLCQ) one should avoid expressions
involving $G_+$, since it is the propagator for the unphysical ("bad")
component of the fermion field that gets eliminated by solving
the constraint equation.
However, since the model that we considered has an underlying Lagrangian
which is parity invariant, one can use $G_+=G_-$ for the self-consistent
solution and still use Eq. (\ref{eq:dmgreen}) or
Eq. (\ref{eq:ctilde}) but with $G_+$ replaced by $G_-$.
In doing so, we obviously used a method to determine
$\Delta m^2_{ZM}$ that is not always applicable. However, the
important point here is not getting the actual numerical
value for $\Delta m^2_{ZM}$ in the above toy model, but 
to demonstrate explicitly that such a value exists which leads to
non-perturbative equivalence between covariant and LF calculations.

\subsubsection{Conclusion:}
There are several things one can learn from this simple toy model.
\begin{itemize}
\item Most importantly, even though the model exhibits spontaneous
breakdown of chiral symmetry, a LF calculation without zero-modes still
gives the same physics (i.e. spectral density) as a covariant calculation.
Thus there is no conflict between trivial vacua on the LF and spontaneous
breakdown of chiral symmetry.
\item The equivalence between LF and covariant approach does not come for free.
It is necessary to introduce an additional renormalization parameter ---
which could however be fixed by imposing conditions derived from parity
invariance.
\item As a byproduct, one can also infer from the above results that the
vertex mass (which multiplies the only chirally odd term in the LF Hamiltonian)
is to be identified with the current quark mass in the covariant approach. This result has been known from perturbative considerations,
but the above example demonstrates its validity in a non-perturbative 
context that even includes spontaneous $\chi$SB as $m\rightarrow 0$.
This result may seem surprising at first since the chirally odd vertex mass
term is also the only term which lifts the degeneracy of the $\pi$ and the
$\rho$ ($J_z=0$). However, hadron wave functions typically have a rather singular
end point behavior in the chiral limit so that matrix elements of the vertex
mass term don't necessarily have to vanish.
\end{itemize}

\subsection{Demise of the Zero-Modes}
Recently, there has been a considerable effort to include explicit
zero-mode degrees of freedom into LF calculations in order
to account for non-trivial vacuum effects (see for example Refs. \cite{pinsky,dave:coral}
and references therein).
There are several comments that I would like to make about these
zero-modes. First, I agree that vacuum effects must have to do
with the $k^+\rightarrow 0$ region. 

However, the examples discussed here and in the references show
that it is not always necessary to include explicit zero-mode degrees
of freedom in order to obtain the right results: with a little extra
effort to properly renormalize, we got away without any explicit
zero-mode degrees of freedom at all.

Would the renormalization be simpler (i.e. less ``independent''
coefficients) if zero-mode degrees of freedom were included? 
In the examples discussed above the answer is no! Including 
explicit zero-modes in the above examples still requires an
infinite kinetic mass counter-term that is not accompanied by
an infinite vertex mass counter-term. This leaves the finite
part of the kinetic mass counter-term ambiguous, i.e. there is 
the same number of renormalization constants.
The root of this perhaps surprising result can be seen best by
approaching LF coordinates using $\varepsilon$ coordinates
\cite{le:ap}. In this approach, LF quantization is defined via
a limiting procedure --- very much like a careful infinite momentum 
boost. If one performs this limiting transition on a finite interval
of length $L$ then the following pattern is observed: 
if one takes the limit $L \rightarrow \infty$ first and the LF limit
next then one gets complete equivalence with the covariant
formulation --- without any non-covariant counter-terms.
However, when one takes the limits in opposite order
(first go to the LF and then $L\rightarrow \infty$)
one still needs to allow for independent renormalization
of vertex mass and kinetic mass --- even if the zero-mode is
included. From a practical point-of-view nothing is thus gained
by including the zero-mode. Including the zero-modes does not
simplify the renormalization procedure.

Going back to the first approach
(with the ordering: $L\rightarrow \infty$ first, LF next)
shows that any nontrivial zero-mode effects do NOT arise from
one single zero-mode, but from an infinite number of modes
in an infinitesimal vicinity of $k^+=0$. It is thus foolish to 
believe that inclusion of LF zero modes (finite intervals on the LF)
will automatically and 
properly take care of the $k^+\rightarrow 0$ physics. Constrained zero modes are not
sufficient to account for all (or even most) aspects of vacuum
structure \cite{dave:coral}.

Since I am criticizing the contemporary zero-mode approaches,
I should also point out the limitations of the conclusions that
one can draw from this work: all the zero modes that played a role
in the examples above were so called {\it constrained zero-modes}
\cite{dave:0}. The proper definition of this terminology can be found
in Ref. \cite{dave:0}, but what it means roughly speaking is zero
modes that have to satisfy some complicated nonlinear constraint
equation, but not a true {\it dynamical} equation of motion.
I don't understand yet what role dynamical zero-modes play on the
LF. However, one result that emerges from this work
is that constrained zero-modes 
can probably \footnote{The reason I included the word ``probably''
here is because I was able to show that they are unnecessary only
by means of examples but not by means of a rigorous proof.}
be omitted completely. Proper renormalization takes
care of them automatically.

Examples for fields that have constrained zero-modes are
scalar fields, fermions and the transverse component of the gauge
fields. One example for fields with dynamical zero-modes is the longitudinal
component $A^+$ of a gauge field.

\section{Summary}
Light-Front Hamiltonians without zero-modes have a trivial
vacuum. Nevertheless, and this is the most important message of these
lectures, nontrivial vacuum structure and trivial LF vacua are not 
contradictions in terms.

Studies in $QCD_{1+1}(N_c\rightarrow \infty)$ showed that spectra
and coupling constants obtained from a LF calculation based on a
trivial LF vacuum still allow to calculate the correct value for the
chiral condensate.
Studies in scalar field theories $1+1$ and $3+1$ dimensions showed
how condensates can appear as explicit parameters in the effective LF 
Hamiltonian. These examples were very helpful to understand why
the apparent conflict between trivial LF vacua and spontaneous
symmetry breaking is no conflict after all. 

The main difference between chiral symmetry breaking and spontaneous 
symmetry breaking in scalar theories is that the order parameter
in scalar fields has the same quantum numbers as the dynamical
degrees of freedom, which facilitates incorporating dynamical
symmetry breaking in the effective Hamiltonian. 
The calculations in a 3+1 dimensional toy model with $\chi$SB, which I
presented in these lectures, were thus very important to confirm
that proper renormalization of LF Hamiltonians leads to agreement
with conventional calculations --- even when chiral symmetry is
spontaneously broken. 

It should be emphasized that the successful LF quantization of the above
1+1 and 3+1 dimensional models has been accomplished without any explicit
zero-mode degrees of freedom. Since the zero-modes degrees of freedom
that would appear in a DLCQ analysis of these models would be so called
{\it constrained zero-modes}, it seems at this point that at least
the constrained zero-modes are unnecessary. 

In all results that I have presented here, it was very important that
the renormalization was properly done. In the models that I mentioned
during this lecture, it seems that we understand now what this means
in practice. However, even though there are now several 3+1 dimensional 
models where this aspect is understood, 
the big remaining challenge is still to construct an ansatz for the
renormalized Hamiltonian for LF-QCD$_{3+1}$.

\noindent
{\bf Acknowledgments}

\noindent
I would like to thank Jim Vary and Frank W\"olz
for the opportunity
to lecture at this school and for their immense efforts that helped
make this school a success. I also want to thank
Hala El-Khozondar, Bob Klindworth and Dave Robertson 
for reading and criticizing the
manuscript. This work was supported by DOE
(grant no. DE-FG03-96ER40965) and by TJNAF.
%

%
%

\end{document}